\documentstyle[aps,epsf]{revtex}

\newcommand{\sign}{\mathop{\mathrm{sign}}\nolimits}
\begin{document}
\title{The continuum gauge field-theory model for low-energy electronic states of icosahedral fullerenes}
\author{D.V. Kolesnikov and V.A. Osipov}

 \address{ Joint Institute for Nuclear Research, Bogoliubov
Laboratory of Theoretical Physics,
141980 Dubna, Moscow region, Russia\\
e-mail: kolesnik@theor.jinr.ru, osipov@theor.jinr.ru}
\date{2 February, 2006}
\maketitle
\begin{abstract} The low-energy electronic structure of icosahedral
fullerenes is studied within the field-theory model. In the field
model, the pentagonal rings in the fullerene are simulated by two
kinds of gauge fields. The first one, non-abelian field, follows
from so-called K spin rotation invariance for the spinor field
while the second one describes the elastic flow due to pentagonal
apical disclinations. For fullerene molecule, these fluxes are
taken into account by introducing an effective field due to
magnetic monopole placed at the center of a sphere. Additionally,
the spherical geometry of the fullerene is incorporated via the
spin connection term. The exact analytical solution of the problem
(both for the eigenfunctions and the energy spectrum) is found.
\end{abstract}

% end of PACS codes

\section{Introduction}

The electronic structure and elementary excitations of \\
fullerene molecules have been of steady interest since the
discovery of the first so-called buckminsterfullerene
C$_{60}$~\cite{kroto} because the knowledge of the electronic
structure gives an important information about the electric and
photo conductivity, magnetic behavior, etc. All these
characteristics were found to be rather unique in fullerene
molecules (see, e.g., review~\cite{gensterblum}), which can be
effectively used in some practical applications in devices based
on fullerenes.

There is a number of different theories to study this problem,
which can be roughly divided into three general groups. The first
one includes empirical methods like the free-electron
gas~\cite{gallup}, and
tight-binding~\cite{manousakis,Lin,Tang,Perez} calculations. The
second one uses {\it ab initio} quantum chemistry
calculations~\cite{dunlap}.
%Notice that both methods
%were found to give the results being in a good agreement with
%experiment.
The third group considers the continuum models within the
effective-mass description~\cite{jose92,jose93,osipov}.

While the continuum description is limited to the electronic
states close to the Fermi level, it has some interesting
attractive features. First of all it gives a possibility to study
large fullerenes where the numerical analysis is a rather
difficult task. Second, it reveals the long-distance physics which
is of importance in various carbon nanoparticles. Finally, the
continuum description allows to elucidate "true" topological
effects like the appearance of the Aharonov-Bohm phase and
anomalous Landau levels due to disclinations.
%~\cite{crespi}.

In this paper, we formulate a continuum model to study low-energy
electron states in icosahedral fullerenes. The model is a variant
of the effective field theory on a sphere describing Dirac wavefunctions
interacting with two types of gauge fluxes. One of the
fluxes is due to so-called $K$ spin rotation invariance
(see~\cite{crespi} for details) and the second one comes from the
local SO(2) invariance of the two-dimensional elastic Lagrangian
in the presence of disclinations~\cite{jpa99}. Actually,
the second flux describes the elastic flow through a surface due to a disclination
and has a topological origin (its circulation is determined by the Frank index,
the topological characteristic of the defect). For this reason, this flux
exists even within the so-called "inextensional" limit (which is usually
adjusted to fullerene molecule~\cite{tersoff}). Notice that the topological
origin the elastic flux results in appearance of the disclination-induced
Aharonov-Bohm-like phase (see~\cite{osipov_pla}).

It should be mentioned that the first continuum model for the
description of fullerene molecule was presented in two
papers~\cite{jose92,jose93}. A different variant of the continuum
model for fullerene was suggested in~\cite{osipov}. The first
model neglected the topological origin of the disclination defects
while the second model missed the $K$ spin rotation invariance.
However, in both papers the non-trivial zero-mode electronic
states are considered. In this paper we found an exact analytical
solution of the problem at low energies.

\section{General formalism}

Let us start from the standard formalism based on the
effective-mass theory proposed in~\cite{mele} to study the
screening of a single intercalant within a graphite host. A
graphite host is considered as a single graphite plane (graphene).
Actually, the effective-mass expansion is equivalent to the
$\vec{k}\cdot\vec p$ expansion of the graphite energy bands about
the $\vec K$ point in the Brillouin zone when the intercalant
potential is zero. In fact, there are two degenerate Bloch
eigenstates at $\vec K$, so that the microscopic wave function can
be approximated by
\begin{eqnarray}
\Psi(\vec k,\vec r)=f_1(\vec k')e^{i\vec k'\vec r}\Psi^S_1(\vec
K,\vec r)+f_2(\vec k')e^{i\vec k' \vec r}\Psi^S_2(\vec K,\vec r)
\label{eq0.0}
\end{eqnarray}
where $\vec k =\vec K+\vec{k'}$. Keeping terms of order $\vec k'$
in the Schr\"{o}dinger equation one can obtain the secular
equation for functions $f_{1,2}(\vec k')$ and after
diagonalization one finally gets the two-dimensional Dirac
equation (see ~\cite{mele} for details)
\begin{equation}
-i\sigma^{\mu}\partial_{\mu}\psi(\vec r)=E\psi(\vec r)
\label{eq:0.1}\end{equation} where $\sigma^\mu$ are the
conventional Pauli matrices ($\mu=1,2$), the energy $E$ is
accounted from the Fermi energy, the Fermi velocity $V_F$ is taken
to be one, and the two-component wave
function $\psi$ represents two graphite sublattices ($A$ and $B$).
As was mentioned in~\cite{mele} the essence of the
$\vec{k}\cdot\vec p$ approximation is to replace the graphite
bands by conical dispersions at the Fermi energy. In addition, one
has to take into account two independent wave vectors ($\vec
K_{+}=\vec{K}$ and $\vec{K}_{-}=-\vec{K}$) in the carbon lattice
which give the same conical dispersion. Therefore, the states
$\mid\vec K_{\pm}A>$ and $\mid\vec K_{\pm}B>$ can be chosen as the
full basis set ~\cite{crespi}.

For our purpose, to take into account both the spherical geometry
of a fullerene molecule and disclinations on its surface  we have
to modify the model (\ref{eq:0.1}) by introducing three additional
fields.

\subsection{The compensating fields}

As a first step, let us consider a single disclination in a
graphene sheet. To introduce the fivefold in the hexagonal lattice
one has to cut the sector $2\pi/6$ and glue the opposite sides
($OB$ and $OB'$ in Fig.1).
\begin{figure}[ht]
\begin{center}
\epsfysize=7cm \epsffile{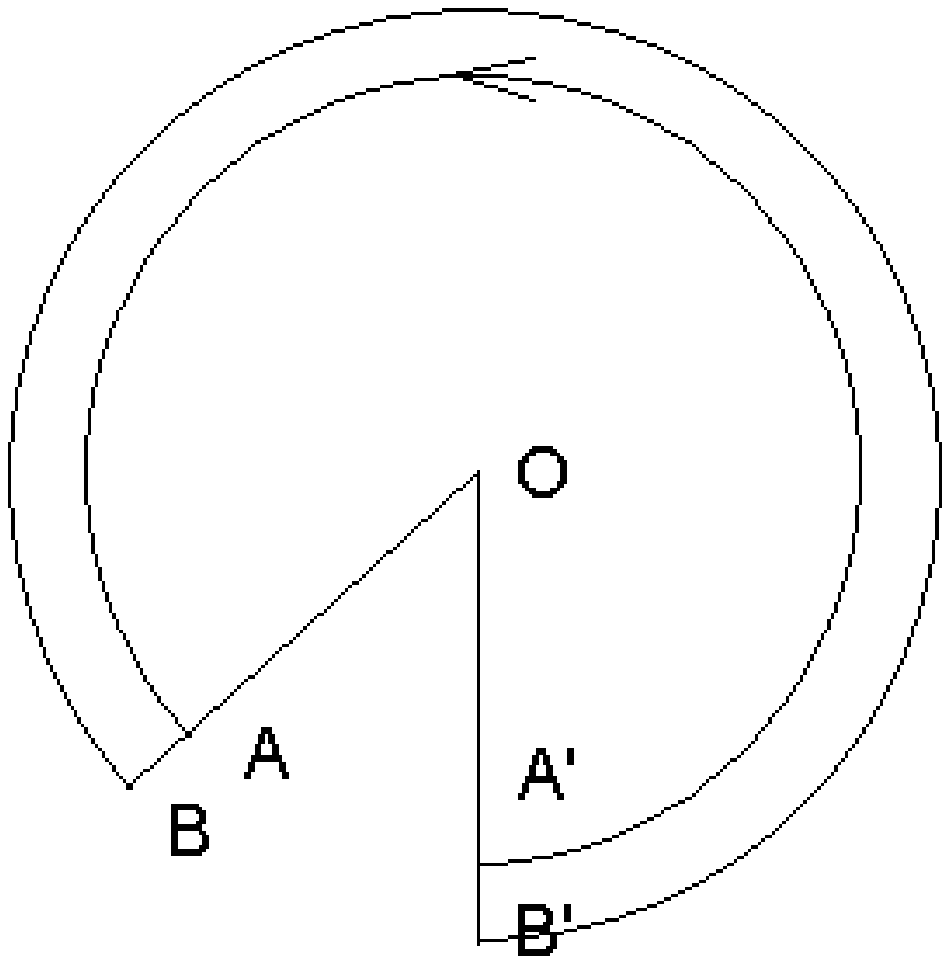}\epsfysize=5cm
\epsffile{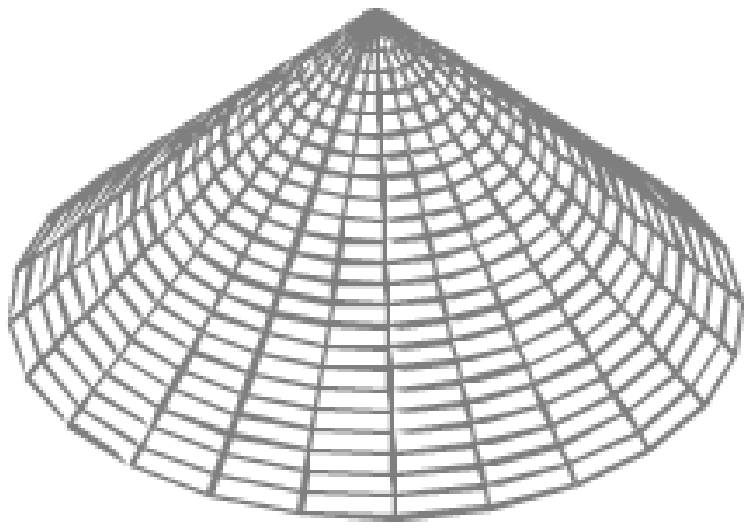} \caption[]{A disclination in elastic membrane
(right) and the image of the membrane (left).}
\end{center}
\end{figure}
\noindent Notice that this is a typical cut-and-glue procedure to
incorporate a disclination in an elastic plane. There exist
several continuum theories for the description of disclinations in
flexible membranes (see~\cite{jpa99,seung,park}) which allow to
formulate similar von Karman equations. In the "inextensional"
limit (free to bend, impossible to stretch), the exact solution
for an isolated positive disclination was found in~\cite{seung}.
It was shown that the elastic membrane becomes buckled and takes a
conical form (see Fig.1). For our purpose, the most appropriate is
the gauge theory of disclinations on fluctuating elastic surfaces
formulated in~\cite{jpa99}. Indeed, this is the most convenient
theory to take into account the topological origin of the
disclination defect. The elastic flux due to pentagonal apical
disclination represented by abelian gauge field $\vec W$ is given
by (see~\cite{osipov,osipov_pla})
\begin{equation}
\oint W_{\mu}dx^{\mu}=\frac{2\pi}{6}\label{Wcirc}.
\end{equation}
As is seen, the elastic flux for pentagonal defect is exactly
$1/6$. It was shown in~\cite{osipov_pla} that the electronic
eigenfunction acquires the additional Aharonov-Bohm-like phase due
to nontrivial topology of elastic surface with a disclination
vortex. In our case, this phase will appear both in the Bloch
(\ref{eq0.0}) and the Dirac (\ref{eq:0.1}) wavefunctions.

To exclude the discontinuous boundary conditions for spinor fields
on the sector sides (lines $OB$ and $OB^\prime$ on Fig.1) let us
introduce two additional fields. First, constructing a boundary
condition for the spinor components  on the line $OB\equiv OB'$
one has to take into account the exchange between A and B
sublattices which takes place for the sector angle $\pi/3$. This
can be done by using an appropriate boundary condition for the K
spin part
%. After the rotation of $K_-$, one can find it
in the form (see~\cite{crespi})
\begin{equation}
\psi(\varphi+2\pi)=-T\psi(\varphi),\quad
T=i\tau_2=\exp(i\frac{\pi}{2}\tau_2)\label{bc}
\end{equation}
where the Pauli matrix $\tau_2$ acts on the K part of the spinor
components, and $\varphi$ is the polar coordinate on the membrane
($0\leq\varphi<2\pi$). One can introduce the non-Abelian gauge
field $\vec{a}$ to eliminate the exponential factor. In this case,
the circulation of the field $\vec{a}$ is written as~\cite{crespi}
\begin{equation}
\oint a_{\mu} dx^{\mu}=\tau_2 \frac{2\pi}{4}\label{acirc}.
\end{equation}
As is seen, this field adds $\pm 1/4$ to the total flux.
%With the field $\vec a$ introduced, the spinor wavefunction in (\ref{bc})
%is changing, as well as $\vec K$ and $\vec{K_-}$, but $\psi^\dag
%\psi$ is, as earlier, the density of states.
It should be
mentioned that the circulation of the gauge field $\vec{a}$ is
governed by topology of the lattice and does not depend upon
geometry of the structure. Thus these operators and fluxes can be
obtained solely from the lattice structure.

The last field to be introduced is the frame rotation field
$\vec{Q}$ which is equivalent to the spin connection
(see~\cite{crespi}). This field (analog to the metrical connection
coefficients) rotates the frame by the angle $2\pi/6$, and its
circulation along the contour $A'A$ is determined by the following
condition:
\begin{equation}
\oint Q_{\mu} dx^{\mu}=-\sigma_3 \frac{\pi}{6}\label{Qcirc}.
\end{equation}
Notice that the spin connection does not contribute to the total
flux.

\subsection{The covariant description and the Dirac equation on
the curved surface}

So far we considered the problem on the plane by using the
discontinuous planar coordinates
%($\vec{R}(R,\phi)$, $0\le \phi<5\pi/3$)
(with the borders of the sector OB and OB$^\prime$). Instead, one
can also use the continuous coordinates on the Riemannian surface
$\vec{r}(r,\varphi)$, $0\le\varphi<2\pi$ (see Fig.1).
 To incorporate fermions on the curved
background we need a set of orthonormal frames $\{e_{\alpha}\}$
which yield the same metric, $g_{\mu\nu}$, related to each other
by the local $SO(2)$ rotation,
$$e_{\alpha}\to e'_{\alpha}={\Lambda}_{\alpha}^{\beta}e_{\beta},\quad
{\Lambda}_{\alpha}^{\beta}\in SO(2).$$ It then follows that
$g_{\mu\nu} = e^{\alpha}_{\mu}e^{\beta}_{\nu} \delta_{\alpha
\beta}$ where $e_{\alpha}^{\mu}$ is the zweibein, with the
orthonormal frame indices being $\alpha,\beta=\{1,2\}$, and
coordinate indices $\mu,\nu=\{1,2\}$. As usual, to ensure that
physical observables are independent of a particular choice of the
zweibein fields, a local $so(2)$-valued gauge field
$\omega_{\mu}$ must be introduced. The gauge field of the local
Lorentz group is known as the spin connection. For the theory to be
self-consistent, the zweibein fields must be chosen to be covariantly
constant~\cite{witten},
$${\cal D}_{\mu}e^{\alpha}_{\nu}=\partial_{\mu}e^{\alpha}_{\nu}
-\Gamma^{\lambda}_{\mu\nu}e^{\alpha}_{\lambda}+(\omega_{\mu})^{\alpha}_{\beta}
e^{\beta}_{\nu}=0,$$
which determines the spin connection coefficients explicitly
\begin{equation}
(\omega_{\mu})^{\alpha\beta}= e_{\nu}^{\alpha}D_{\mu}e^{\beta\nu}.
\label{eq:4}
\end{equation}
Finally, the Dirac equation (2) on a surface $\Sigma$ in presence
of the $U(1)$ external gauge field $W_{\mu}$ and the gauge field
$a_\mu$ is written as
\begin{equation}
i\gamma^{\alpha}e_{\alpha}^{\ \mu}(\nabla_{\mu} -
ia_{\mu}-iW_{\mu})\psi=E\psi, \label{eq:5}
\end{equation}
where
$\nabla_{\mu}=\partial_{\mu}+\Omega_{\mu}$
with
\begin{equation}
\Omega_{\mu}=\frac{1}{8}\omega^{\alpha\ \beta}_{\ \mu}
[\gamma_{\alpha},\gamma_{\beta}]
\label{eq:6}
\end{equation} being
the spin connection term in the spinor representation. Notice that
$\Omega_{\mu}=-i {Q}_{\mu}$ which justifies the above-mentioned
relation between the frame rotation field and the spin connection.
The spinor in (\ref{eq:5}) has the form $ \psi=( F_A^{K} \;
F_B^{K} \; F_A^{K_-} \; F_B^{K_-} )^T $ where $F(\vec r)$ are
envelope functions, \begin{eqnarray}\gamma_{\alpha}=-I\,
\sigma_{\alpha}=- \left(\begin{array}{cc}
  \sigma_{\alpha} & 0 \nonumber\\
  0 & \sigma_{\alpha}
\end{array}\right),\;
 \oint a_{\mu}dx^{\mu}=\frac{2\pi}{4}\tau_{2}I=\frac{2\pi}{4}\left(\begin{array}{cc}
0 & -i\,I\\  i\,I & 0\end{array}\right). \end{eqnarray} The matrix
$\tau_2$ acting in K-spin space appears in (\ref{eq:5}) only
through $a_{\mu}$. Therefore (\ref{eq:5}) can be easily
diagonalized and we arrive at the two-component Dirac equations in
the form
\begin{equation}
-i\sigma^{\alpha}e_{\alpha}^{\ \mu}(\nabla_{\mu} -
ia_{\mu}^k-iW_{\mu})\psi^k = E\psi^k. \label{eq:5a}
\end{equation}
As is seen, the coupled pair of equations (\ref{eq:5}) is reduced to
the decoupled one describing "K-spin up" ($K^{\uparrow}$)
and "K-spin down" ($K^{\downarrow}$) states.
The field $a_{\mu}^k$ is determined by a condition
$$ \oint a_{\mu}^k dx^{\mu}=\pm \frac{2\pi}{4},
$$
with the sign plus (minus) taken for $k=K^{\uparrow}$
($k=K^{\downarrow}$), respectively. Notice that after diagonalization
the four-component spinor $\psi$ is found to be decomposed into
"upper" and "lower" doublet components, $\psi^{K^{\uparrow}}$ and
$\psi^{K^{\downarrow}}$, each of them transforms via SU(2).

\subsection{The continuum model for the icosahedral fullerene}

According to the Euler's theorem, the fullerene molecule consists
of exactly twelve disclinations. Generally, it is difficult to
take into account properly all the disclinations. There are two
ways to simplify the problem. First, one can consider a situation
near a single defect (similar to~\cite{osipov}) taking into
account that each defect in the fullerene can be simulated by two
fluxes: K spin flux (\ref{acirc})  and the elastic flux
(\ref{Wcirc}). In the case of sphere, however, the most
appropriate approximation is to introduce the effective field
replacing the fields of twelve disclinations by the field of the
magnetic 't Hooft-Polyakov monopole with a constant flux density and
the half-integer charge $A$~\cite{jose93}.

The total flux of the monopole $4\pi A$ is equal to the sum of
fluxes from all the disclinations. The procedure of summing up
non-abelian fictitious fluxes from apical defects placed at
different points of the graphite cones was presented
in~\cite{crespi} (so-called $"n-m"$ combination rule).
Namely, for the field $\vec a$ a linear integral circulating any
even number of defects on the lattice is determined by
$\tau_3\, (N\pi/2-(2\pi/3)[(n-m)\,$ (mod $3)])$, where $N$ is a number of defects,
n and m are numbers of steps in positive and negative directions,
respectively (see~\cite{crespi}). The directions rotated by $2\pi/3$ are considered
to be identical. This gives a natural $(n,m)$ classification of two-pentagon
lattices: those for which $n\equiv m$ (mod 3), and those for which
$n\neq m$ (mod 3). This approach is suitable for the case of a sphere.

As is shown below, the eigenfunctions of the low-energy levels
oscillate not too fast with a distance, therefore the effective
"monopole-like" approximation is valid for small quantum numbers
(and near the Fermi energy). So, one can introduce the continuous
field created by the 't Hooft-Polyakov monopole placed at the
center of the sphere. In this case, the contour of integration is
represented by two circles around the poles of the sphere as shown
in Fig.2.
\begin{figure}[!ht]
\begin{center}
\epsfysize=5cm \epsffile{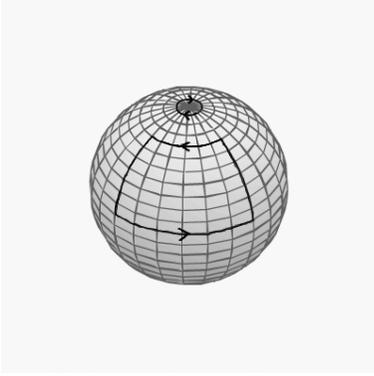} \caption[]{The integration path
on the sphere. The path shown at the center encircles a few
fivefolds. The "total" path of the integration, encircling all
twelve pentagons, includes two contours around the poles:
negatively directed for the north pole (shown) and positively
directed for the south pole (not shown). }
%\label{DOS123}
\end{center}
\end{figure}
\noindent Notice that for fullerenes with the full icosahedral
symmetry (Ih) the combined flux turns out to be a sum of fluxes
due to any pair of defects. Thus, the combined flux does not
depend upon the arrangement of pentagons because any corresponding
fragment of the lattice turns out to be of the above-mentioned
class $(1,1)$. In other words, in icosahedral fullerenes
$(n-m)$\, (mod 3) = 0 due to the mirror symmetry of the
lattice. Finally, the continuous fields take
the form
$$
a^k_{\theta}=0,a^k_{\varphi}=\pm\frac{3}{2}\cos\theta,\quad
W_{\theta}=0,W_{\varphi}=-\cos\theta .
$$
It should be noted that for the Dirac field in the external
potential provided by a monopole, the effective "charge" $A$
involves the "isospin" matrix $\tau'$ (see~\cite{jackiw}). This
matrix can be also diagonalized in the Dirac equation thus giving
the additional sign $\pm$ to the whole charge. What is important, the
coordinate behavior of this field is the same as for the spin
connection field (cf. \cite{abrikosov,jose93}). This fact allows
us to find the exact analytical solution of the problem. Another
significant fact is the presence of "isospin" matrix in the
momentum operator. Similarly to \cite{jose93}
\begin{equation}
J_z=-i(\nabla_{\varphi}-iA{\tau}_{2}'\cos\theta)+\frac{{\sigma}_z}{2}\cos\theta+A\cos\theta{\tau}_{2}'
\end{equation}
and for half-integer $A$ this operator takes \textit{integer} eigenvalues $j=0,\pm 1,\pm
2,...$.

\section{The electronic states of the fullerene}

In accordance with the results of the previous section, the total "charge" is
written as $A=\pm(a^k_{\varphi}+W_{\varphi})/\cos\theta=\pm
1/2,\pm 5/2$. Therefore, the Dirac operator in (\ref{eq:5}) takes
the following form:
$$
\hat{\cal{D}}=-i\sigma_{x}(\partial_{\theta}+\frac{\cot{\theta}}{2})
-i\frac{\sigma_{y}}{\sin\theta}(\partial_{\varphi}-iA\cos\theta).
$$
The substitution
$$
\left(%
\begin{array}{c}
  \psi_{A} \\
  \psi_B \\
\end{array}%
\right) =\sum_j \frac{e^{ij\varphi}}{\sqrt{2\pi}}\left(%
\begin{array}{c}
  u_j(\theta) \\
  v_j(\theta) \\
\end{array}%
\right) ,j=0,\pm 1,\pm 2,\ldots
$$
leads to the equations for $u_j$ and $v_j$
\begin{eqnarray}
-i(\partial_{\theta}+[\frac{1}{2}-A]\cot\theta+\frac{j}{\sin\theta})v_j(\theta)=E
u_j(\theta),\nonumber\\
-i(\partial_{\theta}+[\frac{1}{2}+A]\cot\theta-\frac{j}{\sin\theta})u_j(\theta)=E
v_j(\theta).\label{equv}
\end{eqnarray}
The square of the Dirac operator reads
\begin{eqnarray}
{\hat{\cal{D}}}^2=-[\sigma_{x}(\partial_{\theta}+\frac{\cot{\theta}}{2})+i\frac{\sigma_y}{\sin\theta}(j-A\cos\theta)]^2=\nonumber\\
=-\frac{1}{\sin\theta}\partial_{\theta}\sin\theta\partial_{\theta}+\frac{1}{4}+
\frac{\frac{1}{4}+j^2 +\sigma_z
A}{\sin^{2}\theta}-\frac{\cot\theta}{\sin\theta}(\sigma_z j+2 j
A)+A^2 \cot^2 \theta.\label{D2}
\end{eqnarray}
Let us write the equation $\hat{\cal{D}}^2\psi=E^2\psi$ by using
the appropriate substitution $x=\cos\theta$. From (\ref{D2}) one
obtains
\begin{eqnarray}
\left [\partial_x(1-x^2)\partial_x-\frac{(j-A x)^2-j\sigma_z
x+\frac{1}{4}+\sigma_z A}{1-x^2}\right ]\left(%
\begin{array}{c}
  u_j(x) \\
  v_j(x) \\
\end{array}%
\right)=-(E^2-\frac{1}{4})\left(%
\begin{array}{c}
  u_j(x) \\
  v_j(x) \\
\end{array}%
\right) \label{a0}
\end{eqnarray}
Taking into account the asymptotic behavior of the spinor
functions, one can use the substitution
$$
\left(%
\begin{array}{c}
  u_j \\
  v_j \\
\end{array}%
\right)
=\left(%
\begin{array}{c}
  (1-x)^{\alpha}(1+x)^{\beta}{\tilde{u}}_j (x) \\
  (1-x)^{\gamma}(1+x)^{\delta}{\tilde v}_j (x) \\
\end{array}%
\right),
$$
where
\begin{eqnarray}
\alpha=\frac{1}{2}\left|j-A-\frac{1}{2}\right|,\quad \gamma=\frac{1}{2}\left|j-A+\frac{1}{2}\right|\nonumber\\
\beta=\frac{1}{2}\left|j+A+\frac{1}{2}\right|,\quad\delta=\frac{1}{2}\left|j+A-\frac{1}{2}\right|
\label{alfa}
\end{eqnarray}
Then the equation (\ref{a0}) for the function ${\tilde u}_j$ takes the form
\begin{eqnarray}
(1-x^2)\partial_{x}^2 {\tilde{u}}_j
+(2(\beta-\alpha)-2(\alpha+\beta+1)x)\partial_x {\tilde{u}}_j
+[-2\alpha\beta-\alpha-\beta-\frac{1}{2}(j^2-A^2+\frac{1}{4}-A)+E^2-\frac{1}{4}]
{\tilde{u}}_j =0. \label{eq:20}
\end{eqnarray}
This is equivalent to the Jacobi equation
\begin{eqnarray}
(1-x^2)y''+({\cal{B}}-{\cal{A}}-({\cal{A}}+{\cal{B}}+2)x)y'+\lambda_n
y=0,\nonumber\\
\lambda_n=n(n+{\cal{A}}+{\cal{B}}+1), \label{eq:21}
\end{eqnarray}
with ${\cal{A}}=2\alpha,\, {\cal{B}}=2\beta$, and $n$ being any
non-negative integer. In view of (\ref{eq:20}) and (\ref{eq:21})
one gets the quantization condition
\begin{eqnarray}
\lambda_n=n(n+2(\alpha+\beta)+1)= -2\alpha\beta-\alpha-\beta
-\frac{1}{2}(j^2-A^2+\frac{1}{4}-A)+E^2-\frac{1}{4}.
\end{eqnarray}
Taking into account (\ref{alfa}), one obtains the energy spectrum
in the form $ E_n^2=(n+\alpha+\beta+1/2)^2-A^2$. When $A=0$, one
has $\alpha=(1/2)|j-1/2|,\, \beta=(1/2)|j+1/2|$ and the energy
spectrum is identical to the one found in~\cite{abrikosov} for the
Riemann sphere without a monopole.

In a similar manner one can study the equation for the function
$\tilde{v}$. This gives another energy spectrum
$E_n^2=(n+\gamma+\delta+\frac{1}{2})^2-A^2$. It should be
mentioned that both solutions (for $\tilde u$ and $\tilde v$)
should satisfy (\ref{equv}). This is possible if the condition
$\alpha+\beta=\gamma+\delta$ holds true and, on the other hand, if
one of the energies in the equations for $u$ and $v$ becomes zero.

Let us consider the first case. The possible values of $j$ are found to be $|j|\geq||A|+1/2|$. For $E>0$
one obtains the spectrum
\begin{equation}
E_n^2=(n+|j|+1/2)^2-A^2,\label{levels} \\
\end{equation}
and the eigenfunctions
\begin{eqnarray}
u_j=C_u(1-x)^{\alpha}(1+x)^{\beta} P_n^{2\alpha,2\beta},\nonumber\\
v_j=C_v(1-x)^{\gamma}(1+x)^{\delta}P_n^{2\gamma,2\delta},
\label{eq:25}
\end{eqnarray}
$P_n$ are the Jacobi polynomials. The unit of energy here is $\hbar V_F/R$
where $R$ is the fullerene radius.
One should note that (\ref{levels}) is not allowed for analysis of
zero-mode states. In particular, the degeneracy of the zero-mode
state can not be calculated using (\ref{levels}). According to
(\ref{equv}), the factors $C_u$ and $C_v$ in (\ref{eq:25}) are
interrelated. Indeed, for $j>0$ one has
\begin{eqnarray}
i[(1-x^2)\partial_x -2\gamma-2\beta x]C_vP_n^{2\gamma,2\beta-1}
=E_n(1+x)C_u
P_n^{2\gamma-1,2\beta},\nonumber\\
i[(1-x^2)\partial_x +2\beta-2\gamma x]C_uP_n^{2\gamma-1,2\beta}
=E_n(1-x)C_v P_n^{2\gamma,2\beta-1}.
\end{eqnarray}
Setting $x=1$ in the first equation and using the definition
$$P_n^{a,b}(1)=\frac{\Gamma(a+n+1)}{\Gamma(a+1)n!},\quad
E_n=\pm\sqrt{(2\gamma+n)(2\beta+n)},
$$
one finally gets for $j>0$
\begin{equation}
C_u=-iC_v\sign{E_n} \left( \frac{j+1/2}{j+1/2-A} \right)
\left(\frac{n+j-A+1/2}{n+j+A+1/2}\right)^{1/2}
\end{equation}
In general case, for arbitrary signs of $E$ and $j$, we obtain
\begin{eqnarray}
C_u=-iC_v\sign {(j E_n)}  \left(%
\frac{|j|+1/2}{|j|+1/2-A\sign j }\right)^{\sign j}\left(
\frac{n+|j|-A+1/2}{n+|j|+A+1/2}%
\right)^{1/2}
\end{eqnarray}

In the second case, the possible values of $j$ and $n$ are
determined by $|j|\leq ||A|-1/2|$, $n=0$. One gets exactly one
zero-mode at fixed $j$ and positive fixed $A$
\begin{equation}
u_0=0,\quad v_0=C_v (1-x)^{\gamma}(1+x)^{\delta},\label{zeromode}
\end{equation}
where the relation $P_0^{2\gamma,2\delta}(x)=const$ is taken into
account. Accordingly, if $A<0$ there exists only one zero-mode
solution $u_0$. Thus, for all possible values of $j$ and all
possible positive values of $A$ there exists exactly six different
zero-mode solutions $v_0$.

It should be noted that both in (\ref{eq:25}) and in (\ref{zeromode})
the replacement $j\rightarrow -j,\, A\rightarrow -A$ is equivalent to
the exchange $u\rightarrow v,\, v\rightarrow u$
(cf.~(\ref{equv})). This means that "isospin up" and "isospin down" components
of the spinor on a sphere turns out to be physically equivalent
up to the redefinition of the quantum number $j$ and a unitary transformation.
Therefore, one can restrict consideration to either component, for instance,
to "isospin up".

%Notice that a similar model was investigated
%for Dirac magnetic monopole in~\cite{Wu}.

From (\ref{levels}) and (\ref{zeromode}) one can calculate the
energy spectrum. The possible "charges" are $A=-1/2,5/2$,
so that the first four levels are (in units of $\hbar V_F /R$) the
following: $E=0, 1.41, 2.45, 3.46$. Their degeneracies are $g=6,
2, 6, 6$, respectively. It is interesting to compare these
results with tight-binding calculations. However, two preliminary
remarks should be done. First, the continuum model is correct for
the low-lying electronic states.
Second, the validity of the effective field approximation for
the description of big fullerenes is not clear yet. In fact,
the essence of this approximation is to take into account
the isotropic part of long-range defect fields.
Probably, for bigger fullerenes both the anisotropic part
of the long-range fields and the influence of the short-range
fields due to single disclinations should be properly involved.
Therefore, the exact values of the energy levels do not
agree well with those presented in~\cite{manousakis,Lin,Tang,Perez}.
At the same time, we can surely verify both
the existence of quasi-zero modes found for spherical fullerenes
in~\cite{manousakis,Lin,Tang,Perez} and their 6-fold degeneracy.
There is also a good qualitative agreement in observed scaling of
the energy gap between the highest occupied and lowest
unoccupied energy levels with the size of the fullerene.
%as well as evolution with size of the eigenvalues very close
%to the Fermi energy presented in~\cite{Perez}.

%Actually, the degeneracies
%become doubled if one takes into account both possible signs of
%$A$. To stress the role of the elastic topological flux, let us
%exclude the field $W_{\mu}$ from all the expressions. In this
%case, we arrive at the model similar to~\cite{jose92,jose93}, with
%$A=3/2$. The first four energy levels and degeneracies are found
%to be $E=0, 2.0, 3.16, 4.24$ and $g=3,2,4,6$. While the energies
%appear to be the same as the ones found from the analytical
%solution~\cite{jose92}, the degeneracies are different due to the
%quantum number $n$ that explicitly enters the exact solution. One
%can see that the spectrum changes markedly when the "topological"
%field $W_{\mu}$ is present.

\section{Conclusion}

In this paper, we have studied the electronic states of the
icosahedral fullerene within the continuum field-theory approach.
The influence of the disclinations is taken into account by
introducing an effective field due to magnetic monopole placed at
the center of a sphere and having a total "charge" $A$. The flux
due to monopole is a sum of two fluxes: (i) the K-spin flux and
(ii) the elastic flux due to nontrivial topology of the surface
with a disclination (the pricked out point on the surface). An
exact analytical solution of the problem is found and the explicit
form of the zero-energy modes as well as of the energy spectrum is
presented.

It should be noted that our approach differs from previous studies
of fullerene molecules within the continuum
models~\cite{jose92,jose93,osipov,pincak}. The effective monopole
field introduced in~\cite{jose92} is identical with the K-spin
field $a_\mu$.
%So, the results of~\cite{jose93} can be reproduced
%in our model by excluding the topological field $W_\mu$.
Notice
that a similar to~\cite{jose92} monopole field was used in the
continuum model of the spheroidal fullerenes~\cite{pincak}.
Neither in~\cite{jose92,jose93} nor in~\cite{pincak} the
analytical solution was found. The presented model differs also
from~\cite{osipov} where the gauge field due to K spin was
neglected and only one disclination on a sphere was described.

Introducing the gauge field due to elastic vortex we obtain some
principally new results. First of all, the effective monopole
charge $A$ takes two different values within the proposed model
($-1/2$ and $5/2$) rather than $\pm 3/2$ for icosahedral fullerenes
in~\cite{jose92}. In turn, this finding affects the energy
spectrum which is a combination of spectra for these two charges.
Besides, the eigenfunctions are characterized by different
from~\cite{jose93} conditions for the momentum $j$.
%which results in different degeneracies of the electronic states.

Notice that the more precise description of the fullerenes
requires inclusion of the electron-phonon interaction. In the
simplest form, the role of this interaction was considered
in~\cite{jose92}. It was shown that the energy levels become
shifted and lose a symmetry around the Fermi level. The similar
effect is expected in our model. An interesting open question is
the electronic structure of other (non-Ih) types of spherical
fullerenes as well as of non-spherical (e.g. elliptical) ones.
Probably, the introduction of the monopole-like fields will be
also a good approximation, at least for the low-energy states.

%the A/B sublattice spinor space with the matrices $\sigma$ acting
%on it, the $K/K_-$ space and the $\tau$-matrices, which give after
%the diagonalization positive and negative sign of the field $\vec a$,
%and the "isospin" space, determining the topological origin of
%the monopole and thus the integer values of the momentum.

\vskip 0.5cm

We would like to acknowledge M. Pudlak and S. Sergeenkov for
useful discussions and comments.
This work has been supported by the Russian Foundation for Basic
Research under grant No. 05-02-17721.

\end{document}